\documentclass[10pt, preprint2]{aastex}

\usepackage[]{natbib}

\begin{document}

\title{More on the narrowing of impact broadened radio recombination lines at high principal quantum number}

\author{M.B. Bell\altaffilmark{1}}

\altaffiltext{1}{Herzberg Institute of Astrophysics,
National Research Council of Canada, 100 Sussex Drive, Ottawa,
ON, Canada K1A 0R6; morley.bell@nrc.gc.ca}

\begin{abstract}

Recently Alexander and Gulyaev have suggested that the apparent decrease in impact broadening of radio recombination lines seen at high principal quantum number $n$ may be a product of the data reduction process, possibly resulting from the presence of noise on the telescope spectra that is not present on the calculated comparison spectra. This is an interesting proposal. However, there are serious problems with their analysis that need to be pointed out. Perhaps the most important of these is the fact that for principal quantum numbers below $n$ = 200, where the widths are not in question, their processed generated profile widths do not fit the widths of the processed lines obtained at the telescope. After processing, the halfwidths of the $generated$ and telescope profiles must agree below $n$ = 200 if we are to believe that the processed generated linewidths above $n$ = 200 are meaningful. Theirs do not. Furthermore, we find that after applying the linewidth reduction factors found by Alexander and Gulyaev for their noise added profiles to our generated profiles to simulate their noise adding effect, the processed widths we obtain still do not come close to explaining the narrowing seen in the telescope lines for $n$ values in the range $200 < n < 250$. It is concluded that what is needed to solve this mystery is a completely new approach using a different observing technique instead of simply a further manipulation of the frequency-switched data.

\end{abstract}


\section{Introduction}

The theory of impact broadening of radio recombination lines was developed by \citet{gri67}. A broader coverage was carried out by \citet{gor02} and radio recombination lines have been a valuable tool for many years for the purpose of studying the physical conditions inside galaxies \citep{gor08}.
Using the frequency switching observing technique we have found evidence that the impact broadened linewidths of hydrogen recombination lines near 6 GHz appear to become much narrower than predicted at high quantum numbers, $n > 200$ \citep{bel97,bel00,bel11a,bel11b}. Although a possible explanation for this was suggested by \citet{oks04}, this was refuted by \citet{gri05}. It was later demonstrated by \citet{bel11b} that when observations at other radio frequencies were taken into account the line narrowing appeared to be correlated with the density of recombination lines in frequency space. More recently \citet{hey12} has obtained results that may eventually provide a theoretical explanation for the discrepancies seen between what the impact broadening theories predict and what has been observed in these high-$n$ Rydberg-Rydberg recombination spectra from Galactic HII regions.

Because the multiple overlap data reduction technique we used requires special processing, it has recently been suggested by \citet{ale11} that the narrowing at high $n$ may be related to the processing. After generating theoretical impact broadened line profiles for Orion, these authors claim for the zone between $\Delta n$ = 11 to 14 (near 6~ GHz) that after noise is added to the generated profiles their processed lines exhibit a line narrowing similar to what was reported by \citet{bel11a} as being a mystery. Here we examine their treatment of the data and find, unfortunately, that there appear to be some discrepancies that make their conclusion much less convincing than it appears to be at first glance.

\begin{figure}
\hspace{-0.8cm}
\vspace{-1.5cm}
\epsscale{1.0}
\includegraphics[width=8.5cm]{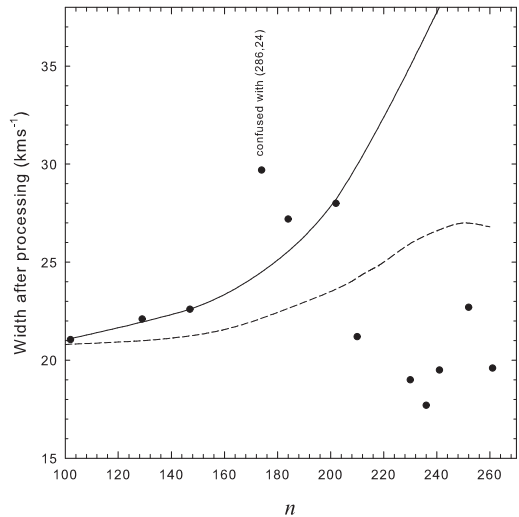}
\caption{{(filled circles)Widths obtained for our telescope lines plotted vs $n$. (solid curve)Widths obtained for our generated Orion Voigt profiles. (dashed curve)Widths of Orion profiles generated by Alexander and Gulyaev.
\label{fig1}}}
\end{figure}

\begin{figure}
\hspace{-0.0cm}
\vspace{-0.5cm}
\epsscale{1.0}
\includegraphics[width=6.5cm]{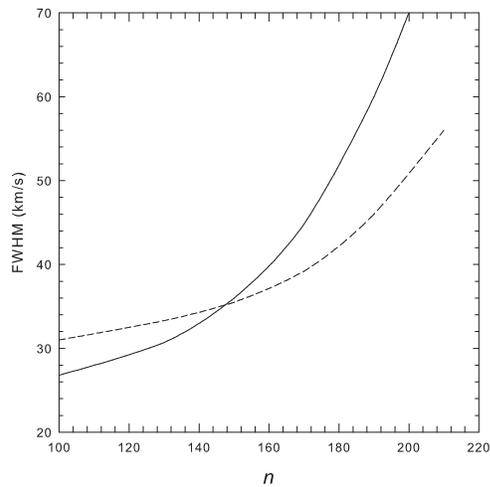}
\caption{{Generated Orion linewidths plotted vs $n$. (solid curve) Widths for our generated Orion Voigt profiles. (dashed line) Impact broadened linewidths calculated by Alexander and Gulyaev for Orion. \label{fig2}}}
\end{figure}

\begin{figure}
\hspace{-0.5cm}
\vspace{-0.5cm}
\epsscale{1.0}
\includegraphics[width=7.5cm]{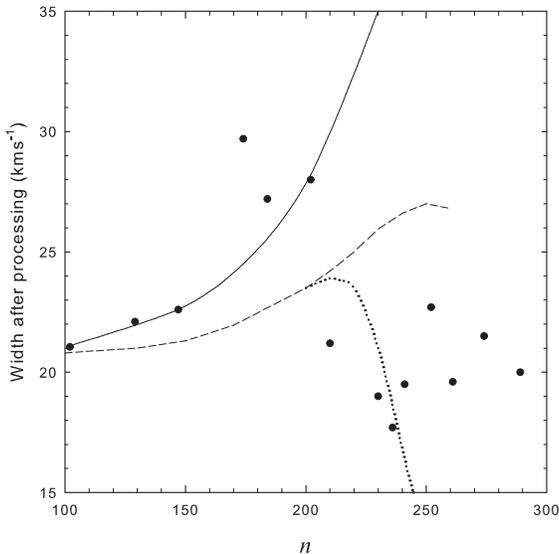}
\caption{{(solid and dashed curves) same as Fig 1. (dotted curve) Processed widths obtained by Alexander and Gulyaev for generated Orion profiles after adding noise. \label{fig3}}}
\end{figure}
 
\begin{figure}
\hspace{-0.4cm}
\vspace{-0.5cm}
\epsscale{1.0}
\includegraphics[width=8.0cm]{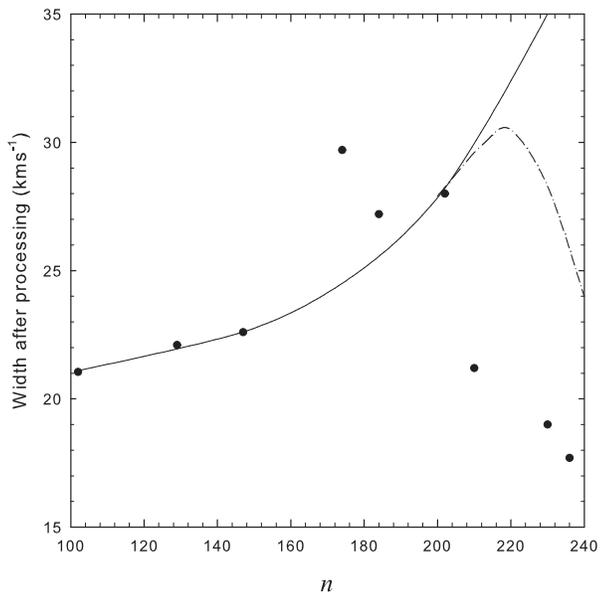}
\caption{{Processed Orion widths plotted vs $n$. (solid curve) same as Fig 1. (dot-dash curve) Widths of our processed profiles after adding noise, obtained using the (noise/no noise) reduction factor found by Alexander and Gulyaev. \label{fig4}}}
\end{figure}

\section{Analysis}

In Fig 1 we have plotted as filled circles the processed linewidths we obtained for the Orion lines observed using the NRAO 140-ft telescope \citep{bel00,bel11a}. We used a frequency-switched offset of $\pm$4 channels which resulted in a separation of 8 channels between signal and reference lines. In processing the data we applied 5 overlaps before cleaning. When frequency switching with multiple overlap reduction is used it is important to realize that the processed widths of lines located at the edge of the observing window cannot be trusted since they can be contaminated by edge effects, especially in strong continuum sources like Orion. For this reason we have previously not included the Orion ($n,\Delta n$) = (194,7) transition in linewidth plots and for this reason it has also not been included here in Fig 1.

In Fig 1 the solid curve represents the processed linewidths obtained for our generated impact broadened Orion Voigt profiles. For transitions below $n$ = 200 the processed linewidths obtained for the generated profiles fit the processed linewidths obtained for the telescope lines remarkably well for the (102,1), (129,2), (147,3), (184,6), and (202,8) lines, all of which are located in reasonably unconfused regions of the spectrum. The width of the (174,5) line is wider than expected and this is assumed to be due to the fact that the (286,24) line is located in its high-frequency wing.

Above $n$ = 200 the widths of the processed telescope lines quickly become much narrower than the widths of the processed generated profiles. One systematic effect that might possibly have explained this line narrowing was the fact that, unlike the telescope profiles, our generated profiles did not contain a superimposed noise component. However, it seemed very unlikely that this would introduce such a rapid narrowing and it was concluded that the Orion lines must become intrinsically narrower than predicted above $n$ = 200.
 
Also plotted in Fig 1 as a dashed curve are the processed linewidths obtained by \citet{ale11} for their generated Orion profiles before adding noise. These two curves should be identical. Clearly they are not, with the dashed curve increasing much more slowly than the solid one. These results can be explained if either, a) the Orion profiles generated by \citet{ale11} are different from the Voigt profiles we generated or, b) their processing has been carried out differently. But there are other problems. Their linewidths, given by the dashed curve, are not a good fit to the telescope data below $n$ = 200 where the widths are not in question. In addition to this, their curve already shows signs of becoming \em narrower \em at high $n$-values, even though there has been no noise added to the generated profiles at this point. 

In Fig 2 the widths of our generated Orion Voigt profiles \em before processing \em are shown by the solid curve. They can be compared to the before processing widths generated by \citet{ale11} shown by the dashed curve. Clearly here as well there is a significant difference between the two curves with the widths of their generated profiles increasing much more slowly than ours. This difference then appears to be the explanation why the two curves in Fig 1 do not agree. However, since the processed widths of our generated curves are in good agreement with the processed telescope lines below $n$ = 200 while theirs are not, we can assume that it is our impact broadened Voigt profiles that have been generated correctly. 

This presents a serious problem. If their processed widths below $n$ = 200 do not fit the widths obtained for the processed telescope lines in this $n$-range how can we believe that the widths they obtained for $n >$ 200 lines are reliable? Furthermore, the processed widths of their generated profiles are already starting to narrow at high $n$ even though there has been no noise added to the spectrum. This indicates that at least some of the narrowing they find at high $n$ is unrelated to the presence of noise on the spectrum and may be due either to the shape of their generated profiles or their reduction process.

In Fig 3 the dotted curve represents the widths obtained by \citet{ale11} for their processed profiles after adding noise to their generated spectra. The solid and dashed curves are the same as in Fig 1. Above $n$ = 220 the noise added widths begin to decrease rapidly and it was concluded by these authors that this explains the decrease seen in the Orion telescope data plotted here as filled circles. This might be a valid conclusion but it would have been much more convincing had their processed generated linewidths agreed more closely with the processed telescope linewidths below $n$ = 200.


Our calculated Orion Voigt profiles were generated by L. W. Avery almost 15 years ago. Unfortunately these profiles, along with our reduction programs, are no longer unavailable to us. Otherwise it would have been easiest simply to add noise to those profiles and reprocess them. In an attempt to obtain an approximate noise-added result for our processed generated profiles, in Fig 4 we have applied the same reduction factor found by \citet{ale11}, between their $before$ and $after$ noise-added curves in Fig 3, to our processed generated widths. The result is shown by the dot-dashed curve in Fig 4. It assumes that all of the narrowing is due to added noise, which may not be the case. The line narrowing can be seen to begin at significantly higher $n$ than does the narrowing in the telescope linewidths and it does not agree well with the widths obtained for the telescope data, especially in the range $200 < n < 240$ ($8 < \Delta n < 14$) where Alexander and Gulyaev agree that noise does not overwhelm the line profiles obtained at the telescope. In fact, in the $n$-region where a fit is claimed by Alexander and Gulyaev (between the vertical bars in their Fig 6) it would appear that the goodness of the fit exists only because the widths, or shapes, of the profiles generated by these authors are incorrect.

\section{Discussion}

The line temperatures were also examined by \citet[see their Fig 5]{ale11}. For RMS noise levels between 0.5 and 1 mK, which was the range we obtained for our telescope data, their line temperature vs $n$ curve falls significantly below our Orion line strengths above $n$ = 225. We also found that the telescope line areas decreased more slowly than the test line areas above ($n,\Delta n$) = (202,8) \citep[see their Figs 3 and 5]{bel00}. It was later pointed out \citep{bel11b} that this could be explained if the widths of these Orion lines were intrinsically narrower than predicted. In fact, this was considered as further proof that the apparent line narrowing above $n$ = 200 was intrinsic and unrelated to the processing. This is related to the fact that in overlapping frequency-switched lines the amount of power lost due to processing decreases as the lines get narrower.

Although these authors have found a possible explanation for line narrowing at high $n$, our previous conclusion is not convincingly altered by their work since, (a) they have no independently observed telescope lines to compare their processed generated lines to, (b) the processed widths of the generated lines they are working with do not fit our telescope data below $n$ = 200, nor (c) above 200 after using their noise/no noise reduction factors, and, d) for 1 mK noise, the temperatures of their processed lines cannot explain the measured line strengths we obtain above $n \sim220$.

These authors argue that the line narrowing we see in the telescope data is the result of "forcing" a program to fit Gaussians to features dominated by noise. Although noise on a line profile can increase the uncertainty in its measured width it is highly unlikely that this could cause such a consistent and rapid narrowing during the Gaussian fitting process when there are as many samples across the profile as here. They could have easily determined whether it was the Gaussian fitting in the presence of a decreasing s/n ratio that caused the narrowing by using their processed generated profiles without noise. By first fitting Gaussians to these profiles and then repeating the fitting after noise was added to the same processed profiles, they could have easily determined whether it was the fitting in the presence of noise that caused the narrowing.

\section{Future Work}

Recent analysis by \citet{hey12} may have some bearing on the discrepancies between what impact broadening theories \citep{gri67,wat06} predict and the narrowing that has been seen in high-$n$ Rydberg-Rydberg recombination spectra from Galactic HII regions. His results suggest that these theories have not allowed for the possibly significant role of the plasma (ion) microfield background in perturbing the radiator states which enter the impact broadening calculations. Violation of the parity selection rule would result in some net diminution per perturber in the effectiveness of the collisions of either type (electron or ion) in "allowed" transitions between sub-states of the same principal quantum number. In spite of  the production of line strengths for 'optically forbidden' transitions, \citet{hey12} finds that an overall loss in collision strength takes place, according to the line strength sum rules, because of appreciable 'self-strength' contributions which would be absent for atomic eigenstates unperturbed by the microfield background. For sufficiently large principal quantum number, a significant departure of line widths from a linear dependence on electron (proton) density may thereby be indicated.
 

Although a theoretical explanation for the apparent line narrowing at high $n$ may eventually be forthcoming \citep{hey12}, the main conclusion that can be drawn from the results reported here is that attempting to solve this line-narrowing mystery using multiple-overlap frequency switching is unlikely to yield conclusive results. It is therefore suggested that using a completely different observing technique may be the best way to address the problem, at least empirically (see for example Roshi et al., 2012). This is especially true when the problem is now reduced simply to one of determining whether the linewidths in question are stronger and narrower than predicted. Stronger, narrower lines are much more easily detected in the presence of baseline structure than are weak, wide lines.

In light of this, the following is suggested as an approach that should give more definitive results for $n$-values above $n$ = 200. Because the predicted, raw impact-broadened linewidths near $n$ = 220 are so large (FWHM $> 90$ km/s), and the widths implied if the line narrowing is real are so much smaller (FWHM $\sim25$ km/s) it should be easy to determine which is correct using position-switching together with a sufficiently long integration time. Selecting a line, or lines, that fall in regions of the spectrum free from other strong confusing lines would be a pre-requisite. Also, position switching between Orion and another strong continuum source that does not contain recombination lines should allow most of the baseline structure to be cancelled. This way questions like how noise on the generated spectrum might affect the result, or whether or not the comparison line profiles have been generated correctly, do not enter the picture.  

\section{Conclusion}

Using a frequency switching and multiple overlap observing technique we have found evidence that above $n$ =200 (near 6 GHz) the widths of impact broadened Hydrogen recombination lines may get much narrower and stronger than predicted. Although Alexander and Gulyaev have claimed that this narrowing occurs because noise was present on the telescope line profiles it has been demonstrated here that because of errors in their analysis their claim may not be valid. If so, and the line narrowing is real, this means that the lines in question should be much easier to detect using more conventional observing techniques than previously thought. Therefore, although a theoretical explanation for the line narrowing may yet be found, for the moment the logical conclusion is that the best way to solve this mystery is to observe one or more of these lines using a different observing technique such as single-dish position switching, or an interferometer. 

\section{Acknowledgements}

I thank J.D. Hey for helpful comments during the preparation of this paper.

\end{document}